# Time dilation and Langevin paradox

**A López-Ramos**
Departamento de Física, Universidad de Oviedo
Edificio Este, Campus de Viesques, 33203 Gijón, Spain

E-mail: lramos@uniovi.es

**Abstract.** The principle of invariance of the velocity of light is only valid for the wrong measurements of inertial observers who ignore their own movement and consider themselves at rest. The Langevin (or clock) paradox arises when it is assumed that the real velocity (and not only the wrong measured velocity) of light in free space can be the same for different observers. Nevertheless, it is generally assumed as true such incoherent assumption that leads to the paradox. To clarify the issue is of great interest to analyse the physics underlying the phenomenon of time dilation (which is not only the result of a mathematical formula) and this has been done in this work. The exact value of light dragging coefficient is also derived.

## 1. Invariance of the velocity of light

We shall begin with the usual example of an inertial observer O' placed at the middle of a closed carriage in a train moving at uniform velocity.

Let us suppose that the observer in the train tries to test his state of movement by sending a light signal to the ends of the carriage, where two mirrors are placed. The carriage does not drag that signal, because the velocity of light in free space does not depend on the velocity of its source. As, during the transmission, the back end of the carriage comes to meet the light signal, while the front end recedes from it; the wave front reaches firstly the mirror in the back end.

Nevertheless, observer O' does not know it yet. He needs to verify the arrivals of the light signals to the ends of the carriage; for instance, by looking at the images of those arrivals. But when coming back, the relative velocities of the signals are exchanged: that coming from the back mirror, has now to pursue observer O'; who, meanwhile, comes to meet the signal proceeding from the front mirror. Therefore, the returning time of each signal is just equal to the going time of the other. The whole go-and-return time is the same for both paths; in such a way that, despite the fact that light reached firstly the back mirror, the observer in the carriage sees both rays of light arriving simultaneously at the carriage ends; and everything happens as if the velocity of light with respect to him *were the same* in the direction of his movement and in the opposite.

## 2. Light partial dragging

The result obtained above is trivial, but observer O' in the carriage can try to test, in a more sophisticated way, whether he is moving, placing the lamp inside a water tank with transparent walls and watching the reflections of the light coming from the back mirrors $B_1$ and $B_2$, and the front mirrors, $F_1$ and $F_2$, respectively, as shown in Fig 1.

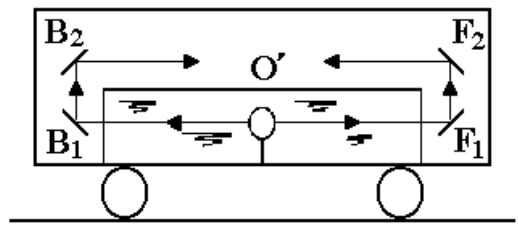

**Figure 1.** Light going through water and returning through air.

Since the velocity of light in water is the quotient between the velocity of light in air and the refractive index of water, the difference in time between the light rays ought to be larger (in absolute value) for the going paths in water than for the returning paths in air. This new attempt, the Mascart-Jamin experiment [1], was made in 1874, by means of the observation of the interference fringes produced by the two rays, using the Earth (which is really moving) as the carriage. But it also fails, as expected by the authors of the experiment: observer O' perceives again the rays simultaneously.

In order to explain a similar experimental result (the maintenance of the aberration angle in telescopes filled with water) and to make the relativity of uniform movement compatible with the undulatory theory of light; Fresnel [2] had predicted, in 1818, that water carries out a partial dragging of light signals.

Let us see, with Mascart, what the Fresnel's dragging coefficient, $f$, (fraction of the velocity of water transmitted to light signals) must be when the velocity of the carriage is much smaller than that of light. Calling $c$ the velocity of light in air; $n$, the refractive index of water, $v$, the velocity of the train and $l$, the carriage length measured by the observer O on the ground; one has, equalizing the time differences in going and returning light paths (Fig. 2):

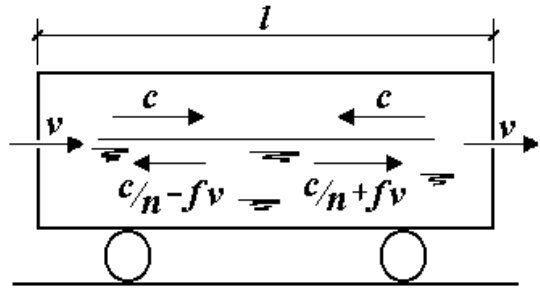

**Figure 2.** Velocities of light signals in water and air.

$$\frac{l/2}{c/n + fv - v} - \frac{l/2}{c/n - fv + v} = \frac{l/2}{c - v} - \frac{l/2}{c + v} \tag{1}$$

from where:

$$\frac{2v(1-f)}{c^2/n^2 - v^2(1-f)^2} = \frac{2v}{c^2 - v^2} \; ; \qquad (1-f)\left(1 - v^2/c^2\right) = \frac{1}{n^2} - \frac{v^2}{c^2}(1-f)^2 \tag{2}$$

and when $v \ll c$:

$$1 - f \approx \frac{1}{n^2} \qquad \Rightarrow \qquad f \approx 1 - \frac{1}{n^2} \tag{3}$$

which is the same value as that obtained by Fresnel (in air $n \approx 1$ and light dragging is negligible).

Fizeau [3] confirmed experimentally, in 1851, Eq. (3), by transmitting light through moving water.

Nevertheless, Fresnel's solution shows weak points. On one hand, it is an "ad hoc" hypothesis to explain an experimental fact; on the other hand, Fresnel had to assume a change of ether density within moving bodies to justify, physically, light partial dragging. And this latter is too little compatible with the fact that the refractive index (and, with it, the change in ether density) depends on light wavelength.

Lorentz [4] gave, in 1892, a theoretical explanation of the light dragging coefficient when building his electron theory.

A simplified deduction can be easily made by using figures 3 and 4. Fig. 3 shows light transmission in quiet water, which consists of the reiteration of two steps: a light transmission in free space from one molecule to another; and a vibration of an electron reached by light.

If we call $l$ the distance between two consecutive molecules of water; $t'_1$, the time needed by light signals for their propagation in free space from one electron to another in the next molecule; and $t'_2$, the time in which signals are an electron oscillation (and consequently they do not move on); the average velocity, $u'$, of light signals in water at rest is

$$u' = \frac{l}{t'_1 + t'_2} = \frac{l}{l/c + t'_2} \tag{4}$$

We obtain the value of $t'_2$ from the refractive index of this medium. Since

$$u' = \frac{c}{n} \tag{5}$$

it results:

$$l = \frac{c}{n}\left(\frac{l}{c} + t'_2\right) \quad \Rightarrow \quad t'_2 = \frac{l}{c}(n-1) = t'_1(n-1) \tag{6}$$

Light transmission when water moves is shown in Fig. 4, where the length contraction of the distance between water molecules is not taken into account, because we are now interested in the non-relativistic case.

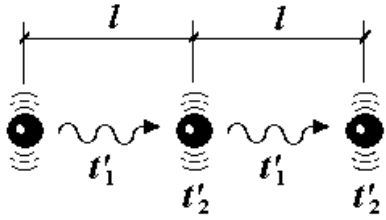 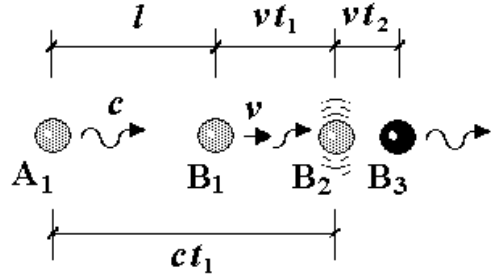

**Figure 3.** Light transmission in quite water.     **Figure 4**. Light transmission in moving water.

The signal sent by molecule A at $A_1$ (being B at $B_1$) reaches molecule B at $B_2$ and is re-emitted by B at $B_3$ position. Thus, the average velocity of light signals in moving water with $v \ll c$ is

$$u = \frac{l + vt_1 + vt_2}{t_1 + t_2} = \frac{l}{t_1 + t_2} + v \tag{7}$$

The time now elapsed in the first step, $t_1$, is obtained from Fig. 4:

$$l + vt_1 = ct_1 \quad \Rightarrow \quad t_1 = \frac{l}{c-v} \tag{8}$$

while $t_2 \approx t'_2$.

We then have, when the velocity of water is much smaller than that of light:

$$u \approx \frac{l}{\frac{l}{c-v} + \frac{l}{c}(n-1)} + v = \frac{c}{\frac{1}{1-v/c} + n - 1} + v \approx \frac{c}{1 + v/c + n - 1} + v$$

$$= \frac{c}{v/c + n} + v = \frac{c}{n(1 + v/nc)} + v \approx \frac{c}{n}\left(1 - \frac{v}{nc}\right) + v = \frac{c}{n} + \left(1 - \frac{1}{n^2}\right)v \tag{9}$$

being in this way theoretically deduced the dragging coefficient.

Lorentz's hypothesis also justifies the change of the refractive index with frequency, as far as the electron oscillation depends on that frequency.

## 3. Time dilation

The theory concerning the principle of relativity of uniform movement was developed when the microscopic constitution of matter was little known. But the mechanism acting in the transmission of light signals through water is rather general. Chemical reactions (including living beings metabolic processes), conscious or unconscious nervous impulses (such as those which control the heartbeats), transmissions of stretchings or displacements in a solid (such as the transmission of centripetal acceleration between next atoms of clock gears), etc., are carried out by means of electromagnetic interactions between electrons and between nuclei and electrons.

Since the velocity of transmission of electromagnetic signals in the space free of matter does not depend on the velocity of their source, signals travelling between particles belonging to a moving body have to pursue the receiving particles, which change their position during the signal transmission. And, although is true that the receiving particle will come towards the signal many times, in cyclic processes the whole time involved increases, as the time saved by particles moving to the signals will not compensate that lost by particles moving away from the signals [5]; just as it happens with the total times of light rays in the well-known Michelson-Morley experiment.

Thus, an immediate consequence of subatomic structure and of the fact that the velocity of electromagnetic waves in free space does not depend on the velocity of their source is that processes in a moving body *run at a slower rate than when it is at rest*. This phenomenon (time dilation) can also be explained with other words: it is due to the fact that moving bodies drag their electrons, but not their photons, so that these last need, in average, more time to go from one electron to another.

In a hypothetical macroscopic system moving with the velocity of light, all of its natural processes would stop. Among all the particles carrying out a physical or chemical process, those placed behind along the direction of movement would not be able to continue the process: electromagnetic signals sent to the other particles, as quick as the signals themselves, would never reach their target. Processes taking place in such a system would only be collisions against external particles or waves found in its way.

It must be emphasized that this phenomenon *is a consequence of classical physics* (unavoidable without a complete dragging of electromagnetic signals). Therefore, it is really surprising to read from an author [6] of several textbooks on relativity that time dilation was a shock for "classical mentality".

The novelty that the principle of relativity introduces about time dilation is that the decrease in the processes rate is not detectable within a system at uniform movement because it is quantitatively unique for all phenomena taking place there. Our train observer cannot realize, for instance, the decrease in his heartbeats rate because it is also the same as the decrease in his clock rate.

This uniqueness is due to two factors: by one side, to length contraction, which equalizes the delay for signals moving parallel or transverse to the direction of movement; and, by another side, to mass increase, which equalizes the delay for all kind of processes.

Larmor [7] was the first, in 1900, who spoke about both time dilation phenomenon and its quantitative uniqueness. Looking for the relationships needed to keep the validity of Maxwell equations with the measurements made by different inertial observers, he found that all internal processes taking place in a system with uniform movement increase their duration according to:

$$\Delta t = \gamma \, \Delta t_0 \qquad (10)$$

where $\Delta t_0$ is their duration when the system is at rest, which is called "proper duration".

This allows us to find the exact value for the velocity of light in moving transparent media, without using the approximation $v \ll c$. In a solid, or in a fluid surrounded by a solid; one has, because of length contraction, that $l = l_0/\gamma$. On the other hand, the electron vibration step also suffers time dilation, due to the increase of inertia, and its duration is not $t'_2$ from Eq. (6), but

$$t_2 = \gamma \, t'_2 = \gamma \, t'_1 (n-1) = \gamma \frac{l_0}{c}(n-1) = \gamma^2 \frac{l}{c}(n-1) \qquad (11)$$

Introducing these corrections into Eq. (9), it results:

$$u = \frac{l}{\frac{l}{c-v} + \gamma^2 \frac{l}{c}(n-1)} + v = \frac{c}{\frac{1}{1-v/c} + \frac{n-1}{1-v^2/c^2}} + v$$

$$= \frac{c(1-v^2/c^2)}{1+v/c+n-1} + v = \frac{c(1-v^2/c^2)+v^2/c+nv}{n+v/c} = \frac{c/n+v}{1+v/nc} \qquad (12)$$

which is the same result as that given by the relativistic formula for the transformation to S of the velocity $c/n$ in S', as it ought to be.

In turn, we can now also deduce the exact value of the dragging coefficient, instead of the approximate value given by Fresnel.
From:

$$u = \frac{c}{n} + fv = \frac{c/n + v}{1 + v/nc} \qquad (13)$$

we have:

$$fv = \frac{c/n + v}{1 + v/nc} - \frac{c}{n} = \frac{c/n + v - c/n - v/n^2}{1 + v/nc} \quad \Rightarrow \quad f = \frac{1 - 1/n^2}{1 + v/nc} \qquad (14)$$

being, finally, this coefficient depending on the value of *v*.

## 4. Langevin paradox

If a spaceship leaves Earth with uniform velocity *v* and changes its velocity to the opposite direction when it is at a distance $l_0$ (measured by the Earth), due to time dilation of moving clocks, there will be a difference

$$\Delta T = 2 \frac{l_0}{v} \left(1 - \sqrt{1 - v^2/c^2}\right) \qquad (15)$$

between the readings of two clocks previously synchronized (at the beginning of the journey), placed on the Earth and on the spaceship, when they meet again (the clock on Earth being ahead).

Langevin or clock paradox arises if the spaceship assumes itself at rest when it is an inertial observer. In such a case it would be the Earth's clock that runs slower both when going and when returning. To accomplish Eq. (15) it is necessary to assume a jump (or an extremely quick variation) in the readings of the Earth's clock just at the moment in which the ship changes the direction of its movement. There would be also a discontinuity in length at that instant: the distance to Earth changes from $l_0/\gamma$ to $l_0$ (when the relative velocity is zero) and again to $l_0/\gamma$, as shown in Fig. 5. The acceleration stages need, then, to be decisive for the solution of the paradox.

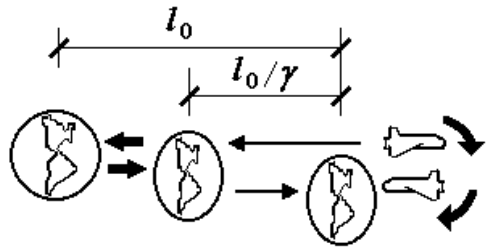

**Figure 5**. Langevin paradox.

But things are really even worse: as some authors [8, 9] remark, the time readings in the Earth's clock would have to drop back if the ship increases its velocity before the reversal of its motion, because this implies that the events happening in the Earth that now become simultaneous for the voyager are in the past of those which were simultaneous before the change of velocity.

What is more paradoxical regarding Langevin paradox is that the facts described in the previous paragraphs are assumed to be true; and it has been said that they are a "consequence of the pseudo-Euclidean geometry of the four-dimensional space-time manifold at variance with the everyday concepts of space and time" [10].

On the contrary, such implausible results come from the incompatible assumptions made by the voyager, of being himself at rest when going and also when returning. If he assumed himself at rest when going, he is obliged to assume as his own velocity when returning the one that he has now in the same inertial frame where he was at rest before.
As the Earth is moving with speed *v* in that frame and the ship returns with a relative velocity *v* in the Earth's measurements, the voyager must now assume that, on account of the formula for the transformation of velocities, his true velocity is:

$$V = \frac{v+v}{1+vv/c^2} = \frac{2v}{1+v^2/c^2} \qquad (16)$$

Thus, when returning, the voyager must consider himself faster than the Earth, deducing that his clock runs slower than the Earth's clock and it must be remarked that the voyager also knows that the velocity of light with respect to him cannot be $c$ again.

The calculations when the spaceship assumes itself at rest during the first stage of the voyage are deduced from Fig. 6.

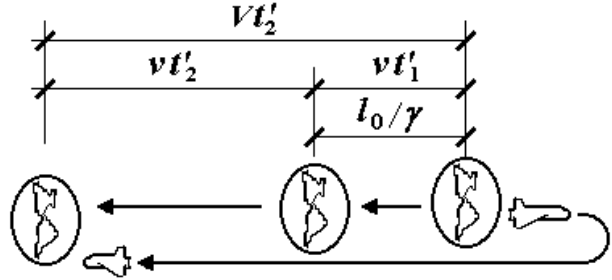

**Figure 6.** Solution of Langevin paradox.

In the first stage, the spaceship is at rest and it will not begin its movement until the distance to Earth is $l_0$ in the Earth's measurements. Due to length contraction of the moving frame where the Earth is at rest, this distance is $l_0/\gamma$ for the ship, and the time elapsed in this stage, $t'_1$ (which coincides with the advance in the reading of the clock on the ship, $t''_1$) is

$$t'_1 = \frac{l_0/\gamma}{v} = t''_1 \qquad (17)$$

while the corresponding advance in the reading of the moving clock on Earth is, on account of time dilation:

$$t_1 = \frac{1}{\gamma}t'_1 = \frac{l_0/\gamma^2}{v} = \frac{l_0}{v}\left(1 - v^2/c^2\right) \qquad (18)$$

In the second stage, the spaceship moves with a velocity $V$ pursuing the Earth, which goes on moving at velocity $v$. Therefore, as deduced from Fig. 6:

$$t'_2 = \frac{l_0/\gamma}{V-v} = \frac{l_0/\gamma}{\frac{2v}{1+v^2/c^2} - v} = \frac{l_0/\gamma}{v - v^3/c^2}\left(1 + v^2/c^2\right) = \frac{l_0}{v\gamma}\frac{1+v^2/c^2}{1-v^2/c^2} \qquad (19)$$

The advance in the reading of the (now moving) clock on the spaceship results:

$$t''_2 = t'_2\sqrt{1-(V/c)^2} = t'_2\sqrt{1-\left(\frac{2v/c}{1+v^2/c^2}\right)^2}$$

$$= t'_2\frac{\sqrt{1+2v^2/c^2+v^4/c^4-4v^2/c^2}}{1+v^2/c^2} = t'_2\frac{1-v^2/c^2}{1+v^2/c^2} = \frac{l_0}{v\gamma} \qquad (20)$$

while the corresponding advance in the reading of the Earth's clock is

$$t_2 = \frac{1}{\gamma}t'_2 = \frac{l_0}{v\gamma^2}\frac{1+v^2/c^2}{1-v^2/c^2} = \frac{l_0}{v}\left(1+v^2/c^2\right) \qquad (21)$$

During the whole journey, the advance in reading of the clock on the ship is given by the sum of Eqs. (17) and (20):

$$t'' = t''_1 + t''_2 = 2\frac{l_0}{v\gamma} = 2\frac{l_0}{v}\sqrt{1 - v^2/c^2} \qquad (22)$$

while that of the Earth's clock is given by the sum of Eqs. (18) and (21):

$$t = t_1 + t_2 = 2\frac{l_0}{v} \qquad (23)$$

Thus, the difference between Eqs. (23) and (22) leads to Eq. (15) without any discontinuity both in length or time, as it must be; and the influence of the acceleration stages is negligible.

Although those last calculations were already included in reference [11], where the final readings of the clocks are deduced for different frames, they are not the calculations given in [11] for the ship frame, which are again those based on the inconsistency of considering that the ship is at rest in the two main stages of its movement.

Of course, as the voyager (who suffers the accelerations) knows perfectly that he is not an inertial observer, he can better correct his measurements from the beginning and accept that his initial reference frame before the journey (the Earth frame) is that in which the velocity of light is isotropic, as has been done in reference [9]. In this case, his calculations are the same as those of observers on Earth.

## 5. Conclusions

When analysing the Langevin paradox, it is necessary to take into account that the voyager changes from an inertial reference frame to another, and that the velocity of light with regard to himself cannot be isotropic twice (when going and when returning). The principle of relativity says that the velocity of light can be assumed as isotropic for *any* inertial reference frame, but *only for one*. To assume that the velocity of light is isotropic for two different inertial frames is a misunderstanding that leads, precisely, to the paradox.

This mistake comes from the so-called "special relativity theory", which rests on a confusion between the results of measurements and the real values of physical magnitudes.

On the other hand, it has been remarked that time dilation is a phenomenon included in classical physics, being its quantitative uniqueness in an inertial moving frame what the principle of relativity adds about it.